\documentclass{jpsj-suppl}
\usepackage{txfonts} 
\usepackage{graphicx}
\usepackage{wrapfig}

\title{Estimate on Spin Asymmetry for Drell-Yan Process at Fermilab with Tensor-Polarized Deuteron}

\author{S. \textsc{Kumano}$^{1,2,3}$ and Qin-Tao  \textsc{Song}$^{1,3}$}

\inst{$^{1}$KEK Theory Center, Institute of Particle and Nuclear Studies, 
KEK, \\
$\ \ $1-1, Oho, Tsukuba, Ibaraki, 305-0801, Japan \\
$^{2}$ J-PARC Branch, KEK Theory Center, Institute of Particle and Nuclear Studies, KEK, \\
$\ \ $and Theory Group, Particle and Nuclear Physics Division,  J-PARC Center, \\
$\ \ $203-1, Shirakata, Tokai, Ibaraki, 319-1106, Japan\\
$^{3}$ Department of Particle and Nuclear Physics, Graduate University for Advanced Studies \\
$\ \ $(SOKENDAI), 1-1, Oho, Tsukuba, Ibaraki, 305-0801, Japan}

\email{qintao@post.kek.jp}

\recdate{October 8, 2016}

\abst{There are four new structure functions for the spin-1 deuteron
in comparison with the ones for the spin-1/2 proton, and they are called 
$b_1$, $b_2$, $b_3$, and $b_4$. The twist-2 structure functions $b_1$ and $b_2$ 
are expressed by tensor-polarized parton distribution functions in the deuteron. 
HERMES measurements of $b_1$ are much different from the prediction of 
the standard deuteron model with D-state admixture. 
It indicates that the structure functions $b_1$ and $b_2$ probe
an interesting new aspect in the deuteron.
There is an approved experiment at JLab to measure $b_1$ and it is expected to start 
in 2019. On the other hand, the measurement of tensor-polarized distributions 
is under consideration at Fermilab by the Drell-Yan process with 
the unpolarized proton beam and tensor-polarized deuteron target.
It is expected to provide crucial information on tensor-polarized antiquark 
distributions. Since the distributions are small quantities,
it is important to estimate the tensor-polarized spin asymmetry 
theoretically to find experimental feasibility 
for an actual proposal at Fermilab.}

\kword{Drell-Yan process, spin asymmetry, tensor structure, quark, gluon, QCD}

\begin{document}
\maketitle

\section{Introduction}
\vspace{-0.05cm}

The quark model was proposed by M. Gell-Mann 
and  G. Zweig in 1964 to classify numerous hadron states, 
and the quark model works successfully to explain major hadron properties. 
In this naive quark model, spins of ground-state hadrons such as the proton 
are carried by quarks. However, the European Muon Collaboration discovered that only
a small fraction of the spin is carried by the quarks in the proton. 
Analyses of subsequent experiments suggest that the fraction is only 20-30\%,
and it is inconsistent with the naive quark-model prediction. 
This spin puzzle of proton suggests that the contributions from
gluon spin and partonic orbital angular momenta should not be zero. 
Therefore, a more sophisticated description is necessary to reveal 
the proton spin structure beyond  the naive quark model. 
In order to solve the spin puzzle of the proton, it is necessary 
to know the gluon-spin and orbital-angular-momentum contributions. 
Recently, generalized parton distributions are widely investigated,
and they provide a possible way to solve the proton puzzle
in the orbital-angular-momentum part.

There is a similar case in the deuteron as the proton spin puzzle.
The deuteron was originally considered as a bound state of proton and neutron 
in S wave. The experimental magnetic moment of deuteron is consistent 
with the S-wave proposal, because the magnetic moment of deuteron is almost 
equal to the sum of magnetic moments of proton and neutron. Later, 
the electric quadrupole moment was observed by experiment, which indicated
that the deuteron cannot be the pure S-wave state. Then the deuteron was 
considered as an S-D mixture state, and the D-wave contribution 
is a very small fraction. This D-state admixture proposal is widely accepted.
However, a HERMES experiment in 2005 \cite{Airapetian:2005cb} showed 
that the twist-2 function $b_1$ is not as small as the prediction 
in the S-D mixture picture \cite{Hoodbhoy:1988am,Jaffe:1988up, Khan:1991qk }. 
The $b_1$ function is expressed by the tensor-polarized 
parton distribution functions (PDFs), 
which could contain the D-wave contribution in deuteron.
However, the standard S-D mixture can not explain the experimental data. 
There are several theoretical studies for the unexpected $b_1$ 
of the deuteron such as six quarks configuration of the deuteron \cite{Miller:2013hla}, 
shadowing effects of the nucleus \cite{Nikolaev:1996jy, Edelmann:1997qe, Bora:1997pi}, 
effects of $\pi$ exchange between proton and neutron \cite{Miller:2013hla}.
At this stage, the structure function $b_1$ and the tensor-polarized PDFs 
are not understood in spite of these theoretical ideas.
In this work, we propose a possible way to clarify tensor-polarized 
antiquark distributions in the Drell-Yan process\cite{Kumano:2016ude},
and it may provide crucial information to find a mechanism for solving 
the issue.

\section{Tensor-Polarized Structure Function $b_1$}
\vspace{-0.05cm}

There are two major methods to investigate the tensor structure functions
of deuteron. 
One is deep inelastic scattering (DIS), and the other is Drell-Yan process.
In the charged-lepton DIS process with the polarized deuteron, there are 
8 structure functions in the hadron tensor
\begin{align}
W_{\mu \nu}^{\lambda_f \lambda_i}=&\frac{1}{4 \pi M} \int  d^4x e^{iqx}   
\left \langle p \, \lambda_f |J_{\mu}(x) J_{\nu}(0)   | p \, \lambda_i  \right \rangle
       \notag \\
   = & -F_1 \hat{g}_{\mu \nu} 
     +\frac{F_2}{M \nu} \hat{p}_\mu \hat{p}_\nu 
     + \frac{ig_1}{\nu}\epsilon_{\mu \nu \lambda \sigma} q^\lambda s^\sigma  
     +\frac{i g_2}{M \nu ^2}\epsilon_{\mu \nu \lambda \sigma} 
      q^\lambda (p \cdot q s^\sigma - s \cdot q p^\sigma )
\notag \\
& 
     -b_1 r_{\mu \nu} 
     + \frac{1}{6} b_2 (s_{\mu \nu} +t_{\mu \nu} +u_{\mu \nu}) 
     + \frac{1}{2} b_3 (s_{\mu \nu} -u_{\mu \nu}) 
     + \frac{1}{2} b_4 (s_{\mu \nu} -t_{\mu \nu}) .
\label{eqn:e1}
\end{align}
Here, $M$, $p$, and $q$ are hadron mass, hadron momentum, and momentum transfer, 
$s^\mu$ is the spin vector of the spin-1 hadron,
$\nu$ is defined by $\nu ={p \cdot q}/{M}$,
and $\epsilon_{\mu \nu \lambda \sigma}$ is an antisymmetric 
tensor with the convention $\epsilon_{0123}=+1$.
The initial and final spin states of the deuteron are denoted
as $\lambda_i$ and $\lambda_f$, respectively.
The notations $\hat{g}_{\mu \nu}$ and $\hat{p}_\mu$ are defined by
$
\hat{g}_{\mu \nu} \equiv  g_{\mu \nu} -{q_\mu q_\nu}/{q^2}, \ \ 
\hat{p}_\mu \equiv p_\mu - ({p \cdot q}/{q^2}) \, q_\mu
$.
The expressions of 
$r_{\mu \nu}$, $s_{\mu \nu}$, $t_{\mu \nu}$, and $u_{\mu \nu}$
are found in Refs. \cite{Hoodbhoy:1988am,Kumano:2014pra}.
The structure functions $F_1$, $F_2$, $g_1$ and $g_2$ exist 
in the hadron tensor of a spin-1/2 hadron as well,
while $b_1$, $b_2$, $b_3$ and $b_4$ are the new
quantities for a spin-1 hadron.

The twist-2 structure functions $b_1$ and $b_2$ are related 
to each other by the relation $2x b_1 =b_2$ in the Bjorken scaling limit.
In the parton picture, $b_1$ has a similar expression as $F_1$,
\begin{align}
F_1= \frac{1}{2 }\sum_i e_i^2 \left [q_i(x,Q^2)+\bar q_i(x,Q^2)   \right ], \ \ 
b_1= \frac{1}{2 }\sum_i e_i^2 \left [ \delta_Tq_i(x,Q^2)+\delta_T \bar q_i(x,Q^2) \right ] ,
\label{eqn:e7}
\end{align}
where $\delta_Tq_i(x,Q^2)=q^0_i-({q_i^{+1}+q_i^{-1}})/{2}$ is
the tensor-polarized PDFs, and $q_i^\lambda$ indicates an unpolarized 
quark distribution with flavor $i$ in the deuteron spin state $\lambda$.
If we integrate $b_1$ by the Bjorken variable $x$, 
we can get a sum rule 
$\int dx \, b_1 (x) = - \lim_{t \to 0} ({5}/{24}) \, t \, F_Q (t) = 0$
\cite{Close:1990zw}
from the valence-quark part;
however, if there are finite $\delta_T \bar q(x)$ distributions,
it becomes 
$
\int dx \, b_1(x)= ({1}/{9}) \! \int \! dx \left[  4  \delta_T \bar u(x)
+  4  \delta_T \bar d(x) \right. $ 
\\ 
\noindent
$\left. +  \delta_T \bar s(x)   \right ]$.
Therefore, a finite sum indicates that there exist 
tensor-polarized distributions for antiquarks in the deuteron.

The deuteron structure function $b_1$ was measured by HERMES Collaboration 
in 2005, and the experimental measurement of $b_1$ is not as small as
the prediction by the standard convolution model with the S-D admixture 
\cite{Airapetian:2005cb}. 
Therefore, a new mechanism could be necessary to explain the tensor structure 
of the deuteron in terms of quark and gluon degrees of freedom.
The integrals of $b_1$ are also provided by the HERMES data as
\begin{align}
\! \! \!
\int _{0.002}^{0.85} \! dx \, b_1(x)&= \left[1.05\pm 0.34(stat) 
      \pm 0.35(sys)  \right ]\times10^{-2}, \ \ 
\left[0.35\pm 0.10(stat) \pm 0.18(sys)  \right ]\times10^{-2} ,
\label{eqn:e3}
\end{align}
where the first value is obtained in the measured range and the second
is in the range of $Q^2 >1$ GeV$^2$.
The nonzero measurements of the $b_1$ integral support the existence 
of tensor-polarized distributions for antiquark quarks $\delta_T \bar q(x)$.
In future, the structure function $b_1$ can be measured at JLab (Thomas Jefferson
National Accelerator Facility), and this should provide us important information
on the tensor structure of deuteron. 

\section{Tensor-Polarized Spin Asymmetry in Proton-Deuteron Drell-Yan Process }
\vspace{-0.05cm}

\begin{wrapfigure}[11]{r}{0.41\textwidth}
   \vspace{-0.1cm}
   \hspace{0.3cm}
     \includegraphics[width=6.0cm]{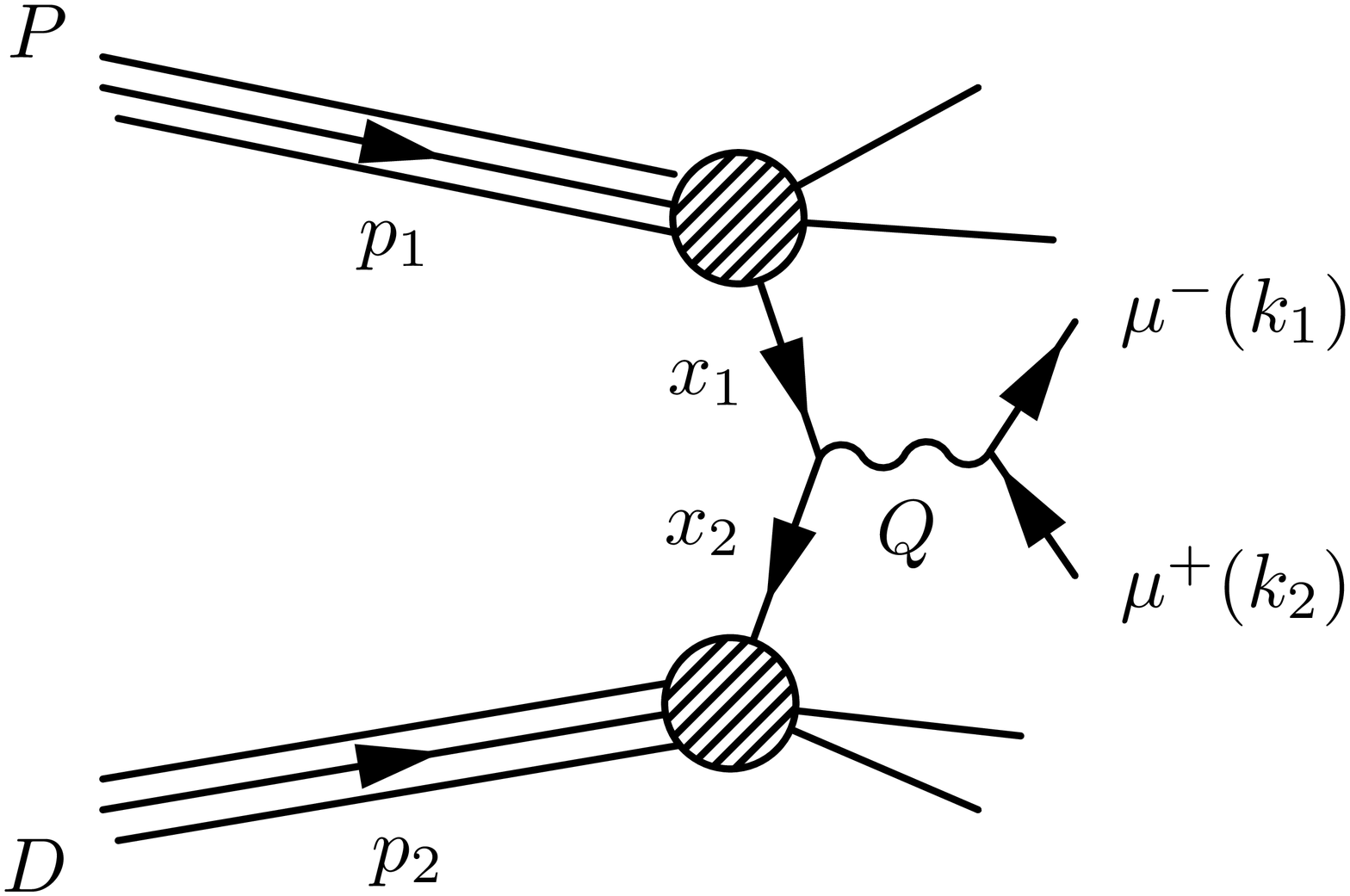}
\vspace{-0.8cm}
    \caption{\label{fig:pd} Proton-deuteron Drell-Yan process.}
\label{fig:gpd-1}
\vspace{-0.5cm}
\end{wrapfigure}

Another possible way to study the tensor structure of the deuteron is 
the Drell-Yan process. It is much better method in probing
$\delta_T \bar q(x)$ directly. 
In the Fermilab E1309 experiment, the beam is unpolarized proton 
(120 GeV, Main Injector) and the target is polarized deuteron.
The proton-deuteron Drell-Yan process 
$ p+d \rightarrow \mu^+ \mu^- + X $
is schematically shown in Fig. \ref{fig:pd}, where
the center-of-mass energy is $s=(p_1+p_2)^2$, 
and the dimuon-mass squared is given by $M^2_{\mu\mu}=Q^2=x_1x_2s$.

There are many spin asymmetries in the proton-deuteron 
Drell-Yan process, but the most important one is new tensor-polarized asymmetry 
$A_Q$ ($\equiv 2 A_{UQ_0}$) defined by
\cite{Hino:1998ww, Hino:1999qi }
\begin{align}
A_{Q}=\frac{1}{\left \langle \sigma  \right \rangle} 
\left [ \sigma(\bullet, 0)-\frac{\sigma(\bullet, +1)+\sigma(\bullet, -1)}{2}  \right],
\label{eqn:e4}
\end{align}
for studying the tensor-polarized PDFs.
Here, $\pm$ and $0$ are for the deuteron spin states
and $\bullet$ indicates the unpolarized proton.
In the leading-order (LO) parton model, it is expressed by the PDFs
\begin{align}
A_{Q}=\frac{\sum_i e_i^2 \left [q_i(x_1) \delta_T \bar q_i(x_2)
+\bar q_i(x_1) \delta_T q_i(x_2) \right]}{\sum_i e_i^2 
\left [q_i(x_1)\bar q_i(x_2)+\bar q_i(x_1)  q_i(x_2) \right]}.
\label{eqn:e4}
\end{align}
A finite spin asymmetry $A_{Q}$ reflects the existence of tensor-polarized
distributions $\delta_T q(x)$ and $\delta_T \bar q(x)$ in the deuteron.
At the large $x_F=x_1-x_2$ region, we have the relations
$q_i(x_1) \delta_T \bar q_i(x_2) \gg \bar q_i(x_1) \delta_T q_i(x_2)$ and
$q_i(x_1) \        \bar q_i(x_2) \gg \bar q_i(x_1)          q_i(x_2)$, 
so that the spin asymmetry of $A_{Q}$ becomes
$
A_{Q}={\sum_i e_i^2 \left [q_i(x_1) \delta_T \bar q_i(x_2) \right]}
/{\sum_i e_i^2 \left [q_i(x_1)\bar q_i(x_2) \right]}
$.
In this case, the existence of  asymmetry $A_{Q}$ at large $x_F$ region 
indicates the tensor-polarized antiquark distributions 
$\delta_T \bar q_i(x)$. The Drell-Yan
process has a merit to measure the antiquark tensor-polarized distributions
directly from the measurement of the asymmetry $A_{Q}$.
In particular, this proton-deuteron Drell-Yan experiment is under consideration
in the Fermilab E1309 experiment as a future project.

\begin{figure}[htpb]
  \centering
  \begin{minipage}[b]{0.45\textwidth}
  \begin{center}
    \includegraphics[width=0.85\textwidth]{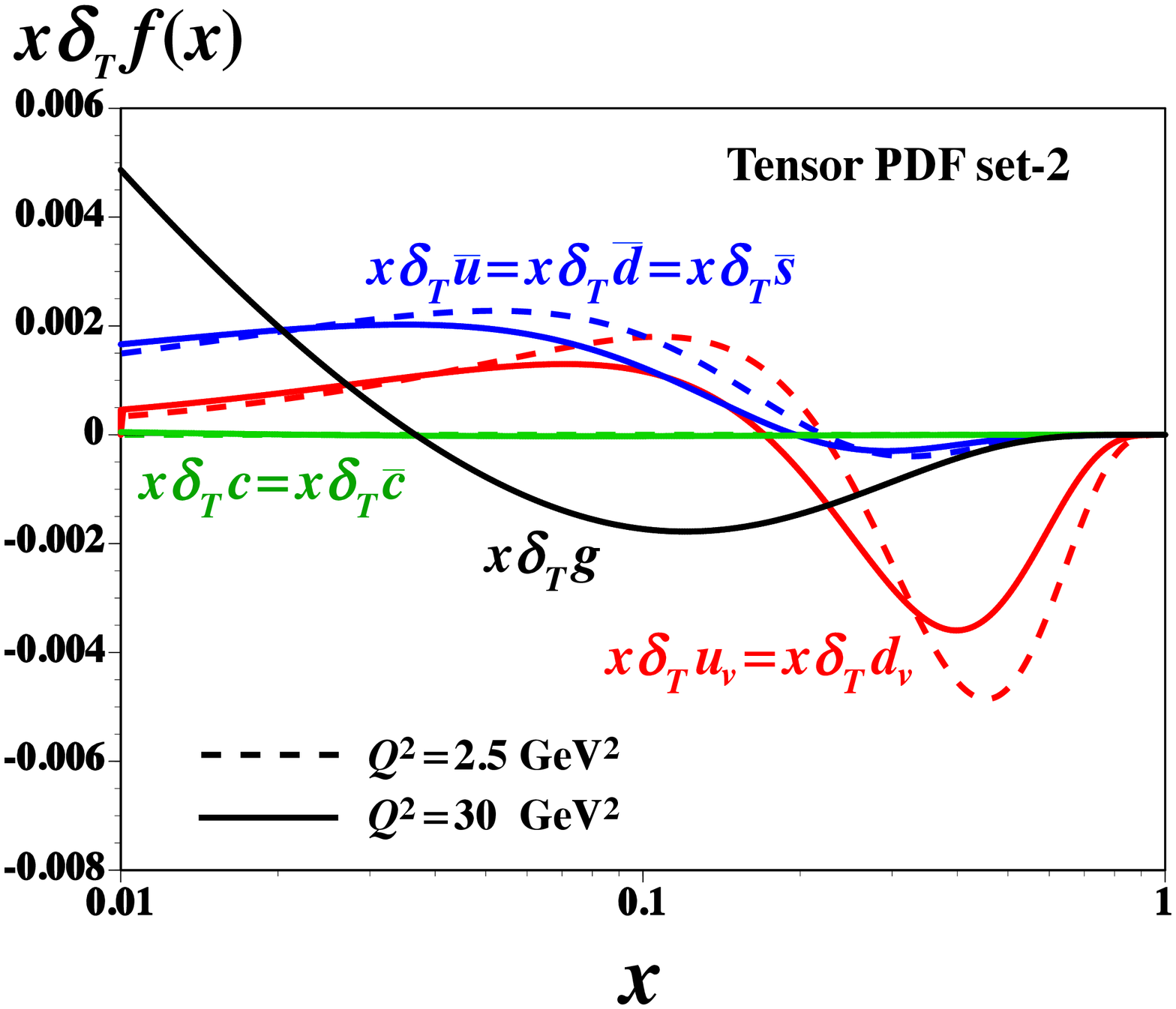}
    \vspace{-0.1cm}
    \caption{\label{fig:tenso-pdfs}Scale evolution of tensor-polarized PDFs of set 2.}
  \end{center}
  \end{minipage}
  \hfill
  \begin{minipage}[b]{0.45\textwidth}
    \includegraphics[width=0.90\textwidth]{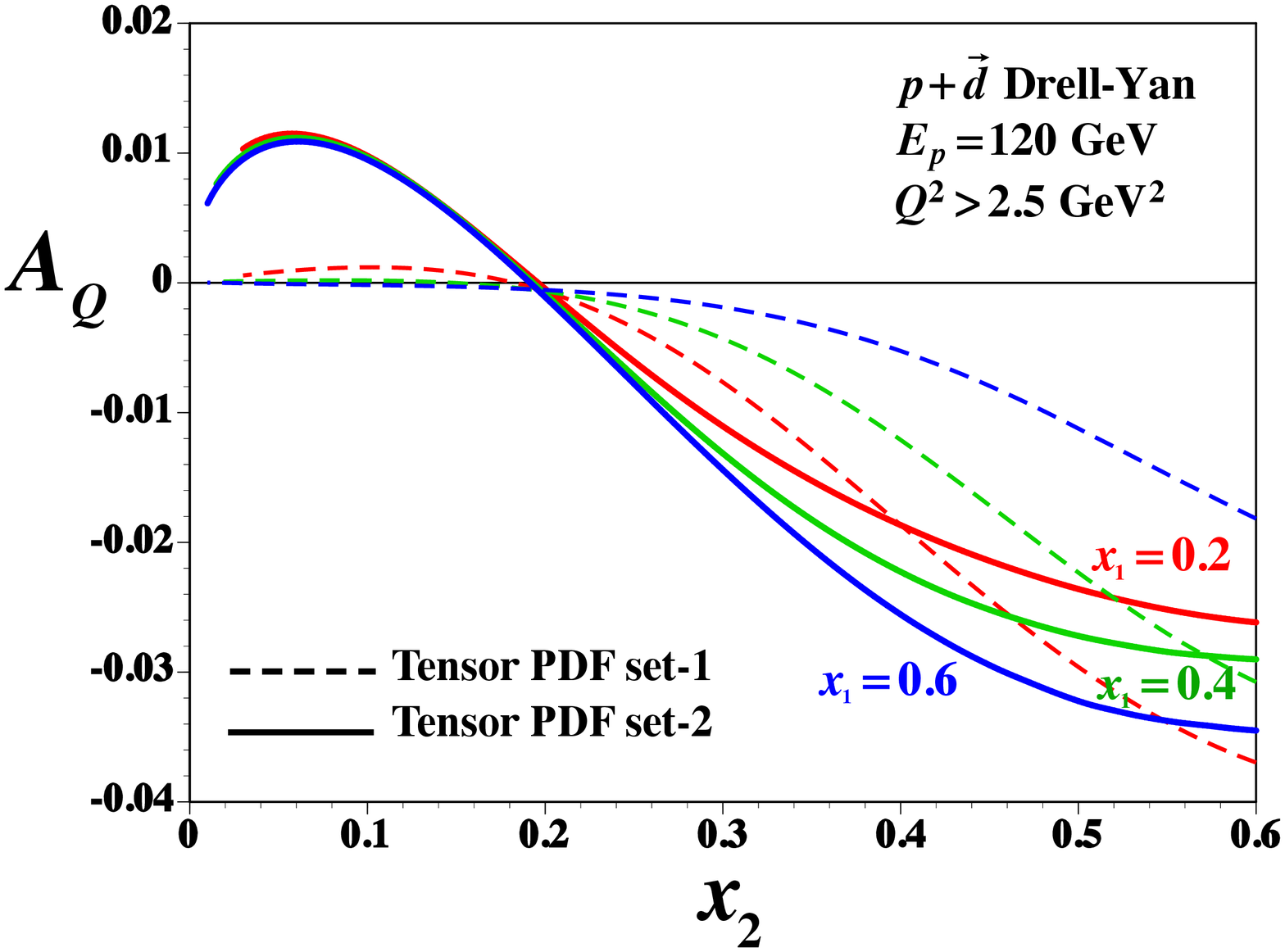}
    \vspace{-0.1cm}
    \caption{ \label{fig:aq} Spin asymmetries are estimated 
    at momentum fractions $x_1=0.2$, $x_1=0.4$ and $x_1=0.6$.}
  \end{minipage}
\vspace{-0.4cm}
\end{figure}

In calculating $A_Q(x_1, x_2)$, the unpolarized distributions of proton 
and deuteron are taken from the MSTW PDFs in the LO \cite{Martin:2009iq}. 
As for the tensor-polarized distributions of deuteron, we use 
the parameterizations of Ref. \cite{Kumano:2010vz} 
for explaining the HERMES data at the average scale $Q^2=2.5$ GeV$^2$. 
There are two analysis types for the parameterizations. 
In set 1, there is no tensor-polarized antiquark distribution 
at $Q^2=2.5$ GeV$^2$. With finite tensor-polarized antiquark distributions,
the set-2 tensor-polarized PDFs can explain the HERMES data better 
than those of set 1. Moreover, the experimental measurements of the $b_1$ 
integral in Eq. (\ref{eqn:e3}) also indicate that the tensor-polarized 
antiquark distributions are necessary at the initial scale. 
By using the DGLAP (Dokshitzer-Gribov-Lipatov-Altarelli-Parisi) evolution 
equations \cite{Hoodbhoy:1988am}, the tensor-polarized distributions 
can also be obtained at larger $Q^2$ scales for the Drell-Yan 
process \cite{Kumano:2016ude}. The evolved tensor-polarized PDFs are shown 
for the set 2 in Fig. \ref{fig:tenso-pdfs} at $Q^2=30$ GeV$^2$ as an example.
It is interesting that we found a finite tensor-polarized gluon distribution
at $Q^2=30$ GeV$^2$, even though they are set to be zero at the initial scale.
The spin asymmetries of set 1 and set 2 are shown in Fig. \ref{fig:aq}
at the momentum fractions $x_1=0.2$, $x_1=0.4$ and $x_1=0.6$ 
\cite{Kumano:2016ude}. The asymmetries of both set 1 and set 2 are a few percent. 
At the small region of $x_2$, the set-1 results are very different 
from those of set 2, because the antiquark tensor-polarized distributions 
become more important in the small $x_2$ region.
The set-2 asymmetries should be more reliable since the tensor-polarized distributions 
can explain the HERMES data well.
Based on our asymmetry estimates, the Fermilab E1309 Collaboration is considering
to measure this tensor-polarization asymmetry.

\section{Summary}
\vspace{-0.05cm}
The tensor-polarized parton distributions can be investigated by
the new structure function $b_1$ and spin asymmetry $A_Q$ in the Drell-Yan process
for the deuteron. They should shed light on a new spin physics, namely
tensor structure in terms of quark and gluon degrees of freedom.
In this work, we showed the theoretical estimates on the spin asymmetry 
$A_Q$, and it is of the order of a few percent. 
In future, those quantities could be measured by Jlab ($b_1$) and 
Fermilab ($A_Q$), and they are expected to clarify the puzzle of deuteron 
tensor structure.

\section*{Acknowledgements}
\vspace{-0.4cm}
This work was partially supported by JSPS KAKENHI Grant Number JP25105010.
Q.-T. S is supported by the MEXT Scholarship for foreign students 
through the Graduate University for Advanced Studies.

\end{document}